\documentclass[twocolumn,showpacs,superscriptaddress,prl]{revtex4}
\usepackage{amssymb, amsbsy, amsmath, latexsym, dsfont, array, layout, graphicx,mathrsfs,color,bm}
\usepackage{xcolor,amsthm,amsfonts}
\usepackage{braket}
\pdfoutput=1
\newcommand{\one}{\mathds{1}}

\begin{document}

\theoremstyle{definition}
\newtheorem{definition}{Definition}

\title{%
    Experimental quantum-walk revival with a time-dependent coin
    }
\author{P. Xue}
\email{gnep.eux@gmail.com} \affiliation{Department of Physics,
Southeast University, Nanjing 211189, China}
\affiliation{%
    State Key Laboratory of Precision Spectroscopy,
    East China Normal University,
    Shanghai 200062, China}
\author{R. Zhang}
\affiliation{%
    Department of Physics, Southeast University, Nanjing 211189, China%
    }
\author{H. Qin}
\affiliation{%
    Department of Physics, Southeast University, Nanjing 211189, China%
    }
\author{X. Zhan}
\affiliation{%
    Department of Physics, Southeast University, Nanjing 211189, China%
    }
\author{Z. H. Bian}
\affiliation{%
    Department of Physics, Southeast University, Nanjing 211189, China%
    }
\author{J. Li}
\affiliation{%
    Department of Physics, Southeast University, Nanjing 211189, China%
    }
\author{Barry C. Sanders}
\affiliation{%
    Hefei National Laboratory for Physical Sciences at Microscale and Department of Modern Physics,
    University of Science and Technology of China, Hefei, Anhui 230026, China%
    }
\affiliation{%
    Shanghai Branch,
    CAS Center for Excellence and Synergetic Innovation Center
        in Quantum Information and Quantum Physics,
    University of Science and Technology of China, Shanghai 201315, China%
    }
\affiliation{%
    Institute for Quantum Science and Technology, University of Calgary, Alberta T2N 1N4, Canada%
    }
\affiliation{%
    Program in Quantum Information Science,
    Canadian Institute for Advanced Research,
    Toronto, Ontario M5G 1Z8, Canada%
    }

\begin{abstract}
We demonstrate a quantum walk with time-dependent coin bias.
With this technique
we realize an experimental single-photon one-dimensional quantum
walk with a linearly-ramped time-dependent coin flip operation and
thereby demonstrate two periodic revivals of the walker distribution.
In our
beam-displacer interferometer, the walk corresponds to movement
between discretely separated transverse modes of the field serving as lattice sites,
and the time-dependent coin flip is effected by implementing a different
angle between the optical axis of half-wave plate and the light propagation at each step.
Each of the quantum-walk steps required to
realize a revival comprises two sequential orthogonal coin-flip
operators, with one coin having constant bias and the other coin
having a time-dependent ramped coin bias, followed by  a conditional
translation of the walker.
\end{abstract}

\maketitle

Quantum walks (QWs) enjoy broad interest due to widespread applications
including to quantum algorithms~\cite{Kem03a}, quantum
computing~\cite{CGW13}, quantum biology~\cite{Llo11}, and quantum simulation~\cite{KRBD10,Kit12,Asb12} plus the fundamental interest of being a natural quantized version of the ubiquitous random walk that appears in statistics, computer science, finance, physics, and chemistry.
QW research has
focused on evolution due to repeated applications of a
time-independent unitary step operator~$U$, but a QW with
time-dependent unitary steps~$U(t)$, with discrete
time~$t\in\mathbb{N}:=\{0,1,2,\dots\}$, opens a much richer array of
phenomena including localization and
quasi-periodicity~\cite{BNP+06,XQTS14}.
Here we demonstrate a
time-dependent QW and use this technique to demonstrate a
revival of the walker's position distribution.

Rather than employing direct time-dependent control, we simulate
time-dependent coin control by setting different coin parameters for
different steps, which are effected in different locations along the
longitudinal axis within our photonic beam-displacer interferometer
(BDI)~\cite{BFL+10}.
The quantum walker within the BDI is a single
heralded photon produced by spontaneous parametric down conversion,
and its walking degree of freedom is the set of discretely spaced
transverse beam modes. The coin flip is effected by employing
quarter- and half-wave plates.

Our method for realizing the first
time-dependent QW
demonstrates the phenomenon of revivals and also
opens the door to realizing a multitude of time-dependent QWs
experimentally.
Compared to prior work employing position-dependent control~\cite{COR+13,SCP+11,KBF+12},
our new technique decreases experimental complexity
by relaxing the requirement of optical compensation.
Our QW revival displays a different characteristic
than typical QW properties such as ballistic spreading and localization of the walker distribution.

The QW with a coin proceeds as a sequence of coin flips and then
walker-coin entangling operations whereby the walker's position is
displaced according to the coin state. We explain the QW now in full
generality so the coin operator admits both spatial and temporal
dependence. Spatially-dependent coin operations have dramatically
demonstrated the realization of topological phases by
QWs~\cite{KRBD10,Kit12,Asb12}, but the time-dependent QW is, until
now, only a theoretical construct and not yet explored
experimentally.

We employ a two-parameter coin realized as a sequence of two rotations,
one with constant bias and another coin that flips along an orthogonal basis
with a time-dependent bias.
This two-parameter coin reveals a richer phenomenological structure than
the case of using just one coin parameter.
One way to picture this richness is to consider the coin operation as being a rotation
of the Bloch sphere, and two parameters allow any rotation to take place whereas one does not.
Thus, having two parameters gives the coin operation full reach over these rotations.

The walker's position on the infinite line is given by the set of
orthogonal states $\{|x\rangle;x\in\mathbb{Z}\}$ ($\mathbb{Z}$ is
the set of all integers), and the homogeneous time-dependent coin
flip $C(t)\in SU(2)/U(1)$ can be expressed as
$C(t)
        =R_x(\Omega t)R_y(\theta)
    =R_x^t(\Omega)R_y(\theta)$
for orthogonal unitary coin-flip operators~\cite{Kem03a}
\begin{equation}
\label{eq:Rxy}
    R_x(\phi)
        =\begin{pmatrix}\cos2\phi &\text{i}\sin2\phi\\
                    \text{i}\sin2\phi & \cos2\phi\end{pmatrix},\,
    R_y(\theta)
        =\begin{pmatrix}\cos2\theta & -\sin2\theta\\
                \sin2\theta & \cos2\theta
        \end{pmatrix}
\end{equation}
with $\phi$ and~$\theta$ the coin biases and~$\Omega$ a constant
pseudo-frequency for ramping one coin bias. The coin eigenstates
are~$\ket{\pm}$.
The unitary conditional
displacement operator for the walker is
\begin{equation*}
\label{eq:FS}
	F:=S\otimes\ket{+}\bra{+}+S^\dagger\otimes\ket{-}\bra{-},\;
	S:=\sum_{x\in\mathbb{Z}}\ket{x+1}\bra{x},
\end{equation*}
and the time-dependent step operator is
$U(t):=FC(t)$.

The Hilbert space for the walker+coin system is
$\mathscr{H}=\text{span}\{\ket{x}\otimes\ket{c}\}$,
and the general state is the density matrix
\begin{equation}
\label{eq:generalestate}
    \rho(t)
    	 =\sum_{x\in\mathbb{Z}}\sum_{c=\pm}\rho_{xc,x'c'}(t)\ket{x}\bra{x'}\otimes\ket{c}\bra{c'}.
\end{equation}
A pure walker-coin system corresponds to $\rho^2=\rho$.
The reduced walker state is the partial trace
\begin{equation}
\label{eq:redwalker}
    \rho^\text{w}(t):=\text{tr}^\text{c}\rho(t)
    =\sum_{x\in\mathbb{Z},c=\pm}\rho_{xc,x'c}(t)\ket{x}\bra{x'}
\end{equation}
with the walker's time-dependent position distribution:
\begin{equation}
\label{eq:pwx}
    p_x^\text{w}(t)=\bra{x}\rho^\text{w}(t)\ket{x}
        =\sum_{c=\pm}\rho_{xc,xc}(t),\,
        \sum_{x\in\mathbb{Z}}p_x^\text{w}(t)=1 \forall t.
\end{equation}
The reduced coin state is $\rho^\text{c}(t)=\text{tr}^\text{w}\rho(t)$.
In our analysis the initial coin-walker
state is a product state with the walker beginning at $\ket{\varphi}^\text{w}:=\ket{x=0}$ and an
initial symmetric coin state
$\ket{\psi}^\text{c}:=\left(\ket{+}+\text{i}\ket{-}\right)/\sqrt{2}$
so $\rho(0)=\ket{\varphi}^\text{w}\!\bra{\varphi}\otimes\ket{\psi}^\text{c}\!\bra{\psi}$.

A QW from time $t=0$ to $t=T$ is effected by the unitary multi-step
operator $\mathcal{U}(T):=\prod_{t=0}^TU(t)$. Our aim is to realize
a time-dependent QW and demonstrate a revival at some time~$T$. We
introduce a rigorous definition of a QW revival here,
motivated by the notion of recurrences in random walks wherein the
walker's position distribution returns to its original distribution
after some time~$T$~\cite{Rev05}.

We define the revival as the return of the walker's position distribution to its original distribution.
The discrepancy between two distributions~$p$ and~$p'$ is quantified by
the total-variation (TV) distance~\cite{SS37}
\begin{equation}
\label{eq:onenormdistance}
	d_\text{TV}\left(p,p'\right)
		:=\frac{1}{2}\|p-p'\|_1
		=\frac{1}{2}\sum_x\left|p_x-p'_x\right|.
\end{equation}
Thus, the discrepancy between the walker's distribution at time~$t$ vs at time~$0$ is
$d_\text{TV}\left(p^\text{w}(t),p^\text{w}(0)\right)$,
which is zero at $t=0$ and at any revival time~$T$.
In the typical case that the initial walker state
is localized at a point, which can be set as the origin,
the mathematics simplifies significantly as we now show.

Under the condition that the walker distribution is localized to a single point,
the distance~(\ref{eq:onenormdistance}) is simply related to the ``QW P\'{o}lya number'',
which is the quantum version of the random-walk P\'{o}lya number
$1-\prod_{t=1}^\infty(1-p^\text{w}_0(t))$.
This P\'{o}lya number is the probability of
the walker ever returning to the origin~\cite{SJK08,SKJ08}.

A QW is ``recurrent'' only if the QW P\'{o}lya number is one
and ``transient'' otherwise~\cite{SJK08}.
The simple relation between the QW P\'{o}lya number and TV distance
for a walker initially localized at $x=0$ (i.e., $p_x^\text{w}(0)=\delta_{x0}$) is
\begin{equation}
\label{eq:dTVp0}
	d_\text{TV}(t)=1-p_0^\text{w}(t),
\end{equation}
which is convenient experimentally as only a projective measurement onto the origin of the walk is required.

A QW revival is achieved at time~$T$ if
$\mathcal{U}(T)=\one\otimes\mathcal{C}(T)$
for some coin operator~$\mathcal{C}(T)$;
i.e., the walker and coin evolve with period~$T$
from separable state
$\ket{\varphi}^\text{w}\!\bra{\varphi}\otimes\ket{\psi}^\text{c}\!\bra{\psi}$
to separable state
$\ket{\varphi}^\text{w}\!\bra{\varphi}\otimes\ket{\psi'}^\text{c}\!\bra{\psi'}$
for $\ket{\psi'}^\text{c}:=\mathcal{C}(T)\ket{\psi}^\text{c}$.

We denote a length~$T$ string of coin flips as the vector $\bm{c}$ whose elements are~$\pm$,
and we denote the balanced subset by~$\mathbb{C}_T$,
which comprises all strings~$\bm{c}$ with an equal number of~$+$ and~$-$ elements.
Consequently~$T$ must be an even number.
Then the effective coin operator for time~$T$ takes the form
\begin{align}
	\mathcal{C}(T)
		=&\sum_{\bm{c}\in\mathbb{C}_T}\ket{c(T)}^\text{c}\!\bra{c(T)}C(T)
			\cdots\ket{c(1)}^\text{c}\!\bra{c(1)}C(1)
				\nonumber\\
		 =&\sum_{\bm{c}\in\mathbb{C}_T}\prod_{t=1}^T\ket{c(t)}^\text{c}\!\bra{c(t)}C(t),
\label{eq:coin}
\end{align}
which yields directly $\left\langle\mathcal{C}^\dagger(T)\mathcal{C}(T)\right\rangle=1$.

As a special case $\mathcal{C}(T)=\mathds{1}$ is possible and corresponds
to a joint revival of both the walker and coin state,
which we call a ``complete revival'' and implies $\ket{\psi'}^\text{c}=\ket{\psi}^\text{c}$.
Experimentally we show a complete revival at~$T$
by demonstrating that the TV distance between the walker distributions at~$t=T$ vs at $t=0$ is small
and we show that both overlaps
$\mathcal{O}:={}^\text{c}\!\bra{\psi}\rho^\text{c}(T)\ket{\psi}^\text{c}$
and
$\mathcal{O}':={}^\text{c}\!\bra{\psi'}\rho^\text{c}(T)\ket{\psi'}^\text{c}$,
calculated from the tomographically reconstructed reduced coin state,
are close to one.

In Table~\ref{table:revival},
we show for each~$T$ up to~$8$ which values of~$\theta$ and~$\Omega$
yield revivals and show in bold which values also give $\mathcal{C}(T)=\mathds{1}$.
Experimentally we realize four of the cases in the table,
corresponding to $T=8$,
thereby demonstrating revivals of a time-dependent QW.
\begin{table}[htbp]
\begin{tabular}{|c||r|c|}
  \hline
  $T$ & $\theta/\pi$ & $\Omega/\pi$ \\ \hline
  $2$ & $0$ & $1/8$,$3/8$ \\ \hline
  $2$ & $1/4$ & $\bm{0}$,$1/4$,$\bm{1/2}$ \\
  \hline
  $4$ & $0$ & $1/12$,$\bm{1/4}$,$5/12$ \\
  \hline
  $4$ & $1/4$ & $\bm{0}$,$1/6$,$1/3$,$\bm{1/2}$
  \\ \hline
  $6$ & $0$ & $1/16$,$3/16$,$5/16$,$7/16$
  \\ \hline
  $6$ & $1/4$ & $\bm{0}$,$1/8$,$\bm{1/6}$,$1/4$,$3/8$,$\bm{1/3}$,$\bm{1/2}$
  \\ \hline
  $8$ & $0$ &\fbox{$1/20$},\fbox{$\bm{1/8}$},$3/20$,$\bm{1/4}$,$7/20$,$\bm{3/8}$,$9/20$
  \\ \hline
  $8$ & $1/4$ & $\bm{0}$,\fbox{$1/10$},\fbox{$\bm{1/8}$},$1/5$,$\bm{1/4}$,$3/10$,$\bm{3/8}$,$2/5$,$\bm{1/2}$ \\
  \hline
\end{tabular}
\caption{%
	Rotation parameters~$\theta$ and~$\Omega$ leading to a revival
	for $T\leq 8$, and bold parameters are cases for which $\mathcal{C}_T=\mathds{1}$,
	hence revivals of both walker and coin for given~$T$.
	The four experimentally realized cases are shown in boxes for $T=8$.%
	}
\label{table:revival}
\end{table}

\begin{figure}
\includegraphics[width=\columnwidth]{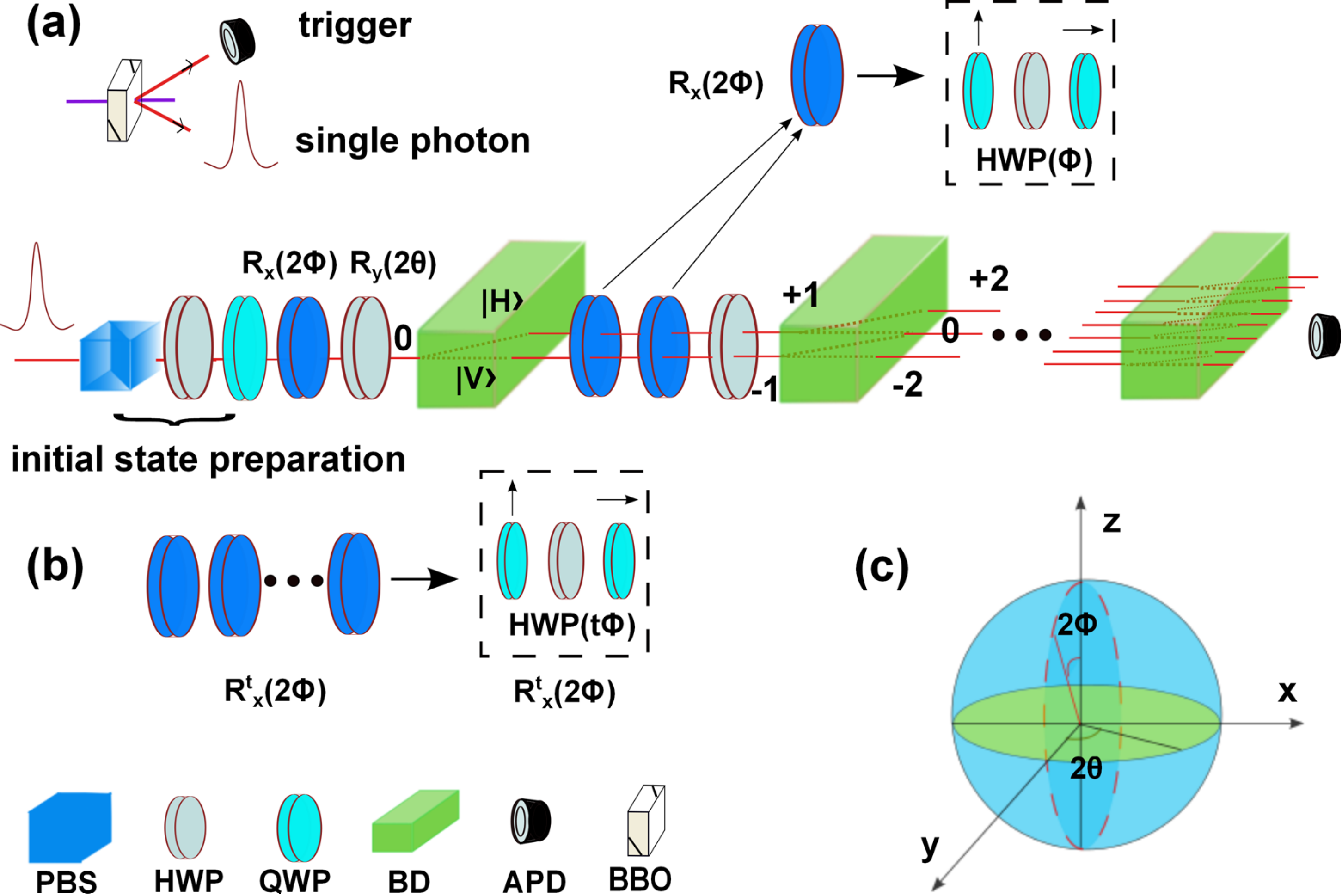}
\caption{
    (Color online)
    Scheme for realizing a $T$-step QW with time-dependent coin
    flipping. (a)~A $\beta$-Barium Borate (BBO) crystal is pumped to yield two correlated photons,
    one arriving at a trigger detector to herald the other photon;
    experimentally this trigger-herald pair is registered by a coincidence count at the
    two avalanche photo diodes (APDs).
    The initial walker-coin state is prepared by the heralded photon
    proceeding through a polarizing beam splitter (PBS)
    and then a half-wave plate (HWP) and a quarter-wave plate (QWP)
    prior to arriving at the first beam displacer (BD).
    A combination of a QWP-HWP-QWP sequence, depicted by a cuboid,
    implements the coin-flip operator $R_x(\Omega)$,
    and the HWP implements $R_y(\theta)$.
    Coin-flip operator $R_y(\theta)$ is realized with a HWP angle set to $\theta$.
    (b)~Implementation of $R_x(\Omega t)$ as a HWP set to $\Omega t$
    $(\bmod 2\pi)$ sandwiched between two QWPs. (c)~Bloch sphere representation of the rotations on
    the coin.}
\label{fig:setup}
\end{figure}

We now describe the experimental realization of the QW with revivals,
which uses an 800.0~nm single heralded photon as the walker.
The detailed scheme is depicted in Fig.~\ref{fig:setup}
and explained fully in the figure caption.

To make the walker,
we generate a wavelength-nondegenerate polarization-degenerate
pair of photons by type-I spontaneous parametric down conversion
in a pair of back-to-back nonlinear-$\beta$-barium-borate (BBO) crystals
pumped by a $400.8$~nm CW diode laser with up to $100$~mW of power;
one photon serves as the trigger detected by the avalanche photo diode (APD) detector
with a dark count rate of $<100$~s$^{-1}$.

The trigger photon heralds the walker photon through two-photon coincidence detection.
More than $20000$ coincidence counts are detected in an overall measurement time. Although our system is operated in continuous-wave mode, photodetection events effectively post-select propagating pulsed-photon states.
The probability for
more than one photon pair is less than $10^{-4}$ hence neglected.
The walker photon goes through a BDI~\cite{BFL+10,XQTS14,XQT14},
which is a robust interferometer that enables more steps than is the case for other interferometers.

The BDI is a sequence of birefringent calcite beam displacers (BDs).
Each BD with length $28.165$~mm splits a beam into two parallel beams,
which are recombined interferometrically at the next BD
with interference visibility reaching $0.996$ per step.
The BDs increase in size as the step number increases,
and the largest BD has a clear aperture 33~mm $\times$ 15~mm.

The optical axis of each BD is cut so that vertically polarized light is
directly transmitted and horizontal light undergoes a 3~mm lateral
displacement into a neighboring mode which interferes with the
vertical light in the same mode.
These parallel beams create an integer lattice perpendicular to the direction of beam propagation.

The initial walker-coin product state is prepared in a product state by directing
the photon through a PBS, HWP, and finally QWP at the initial phase of the BDI;
these three components control the polarization state of the walker photon.
The state~$\ket{+}$ corresponds to a single horizontally polarized photon~$\ket{H}$
and the state~$\ket{-}$ to a single vertically polarized photon~$\ket{V}$.
Interference filters restrict the
photon bandwidth to $3$~nm, and these bandwidth-limited photons are then
steered into the optical modes of the linear-optical network formed
by a series of BDs.

Subsequent to preparing the initial walker-coin state,
each QW step proceeds by directing the beam sequentially through regions
comprising a BD followed by a HWP to effect~$R_y(\theta)$ with~$\theta$
the HWP tilt angle and finally through a second HWP sandwiched between two QWPs.
The two QWPs have  optical axes
oriented vertically and horizontally respectively,
and the second HWP is set at $\Omega t\mod 2\pi$.

Together the HWP and the sandwiching QWPs collaborate to effect the unitary operator $R_x(\Omega t)$. Compared to the previous work by using position-dependent phase shifters~\cite{XQTS14,XQT14}, the sandwiched waveplates are an important technical advance for QW interferometry because the sandwiched waveplates do not require optical compensators whereas position-dependent phase shifters cause photons to experience phase differences due to optical path differences and must be compensated.
The walker's evolution time~$t$ thus corresponds to longitudinally sequential QW steps.
With this scheme
we are able to realize eight steps with the number of steps

We now present in Table~\ref{table:data} the results of our experiments for four parameter choices
of $\theta$ and~$\Omega$
given in the boxes of Table~\ref{table:revival}.
In all cases the small distances $d_\text{TV}$ indicates excellent QW revivals,
and overlaps $\mathcal{O}'$ close to $1$ indicate that the coin state is quite close to the theoretical state.
\begin{table}
\begin{tabular}{|c|c|c|c|c|c|c|c|c|}
  \hline
	$N$ &$\theta/\pi$ & $\Omega/\pi$ & $d_\text{TV}$ & $\Delta d_\text{TV}$
	& $\mathcal O$ & $\Delta\mathcal O$ & $\mathcal{O}'$ & $\Delta\mathcal{O}'$\\
  \hline\hline
  $16$ &$0$ & $1/8$ & $0.197$ & $0.043$ & $0.818$ & $0.024$ & $0.818$ & $0.024$\\
  \hline
  $16$ &$1/4$ & $1/8$ & $0.206$ & $0.039$ & $0.803$ & $0.025$ & $0.803$ & $0.025$\\
  \hline
  $8$ & $0$ & $1/20$ & $0.109$ & $0.020$ & $0.642$ & $0.018$ & $0.972$ & $0.019$\\
  \hline
  $8$ & $1/4$ & $1/10$ & $0.095$ & $0.015$ & $0.640$ & $0.025$ & $0.971$ & $0.023$\\
  \hline
\end{tabular}
\caption{%
	For four choices of~$\theta$ and~$\Omega$ and measuring at the first (or second) revival,
	experimental results for TV distances obtained from
	measured $p_0^\text{w}$ via Eq.~(\ref{eq:dTVp0}),
	overlap~$\mathcal{O}$ between initial and final coin state,
	overlap~$\mathcal{O}'$ between final coin state and theoretically predicted coin state.
	Error bars are in columns indicated with~$\Delta$. $N$ is the number of the steps.%
	}
\label{table:data}
\end{table}

\begin{figure}
\includegraphics[width=\columnwidth]{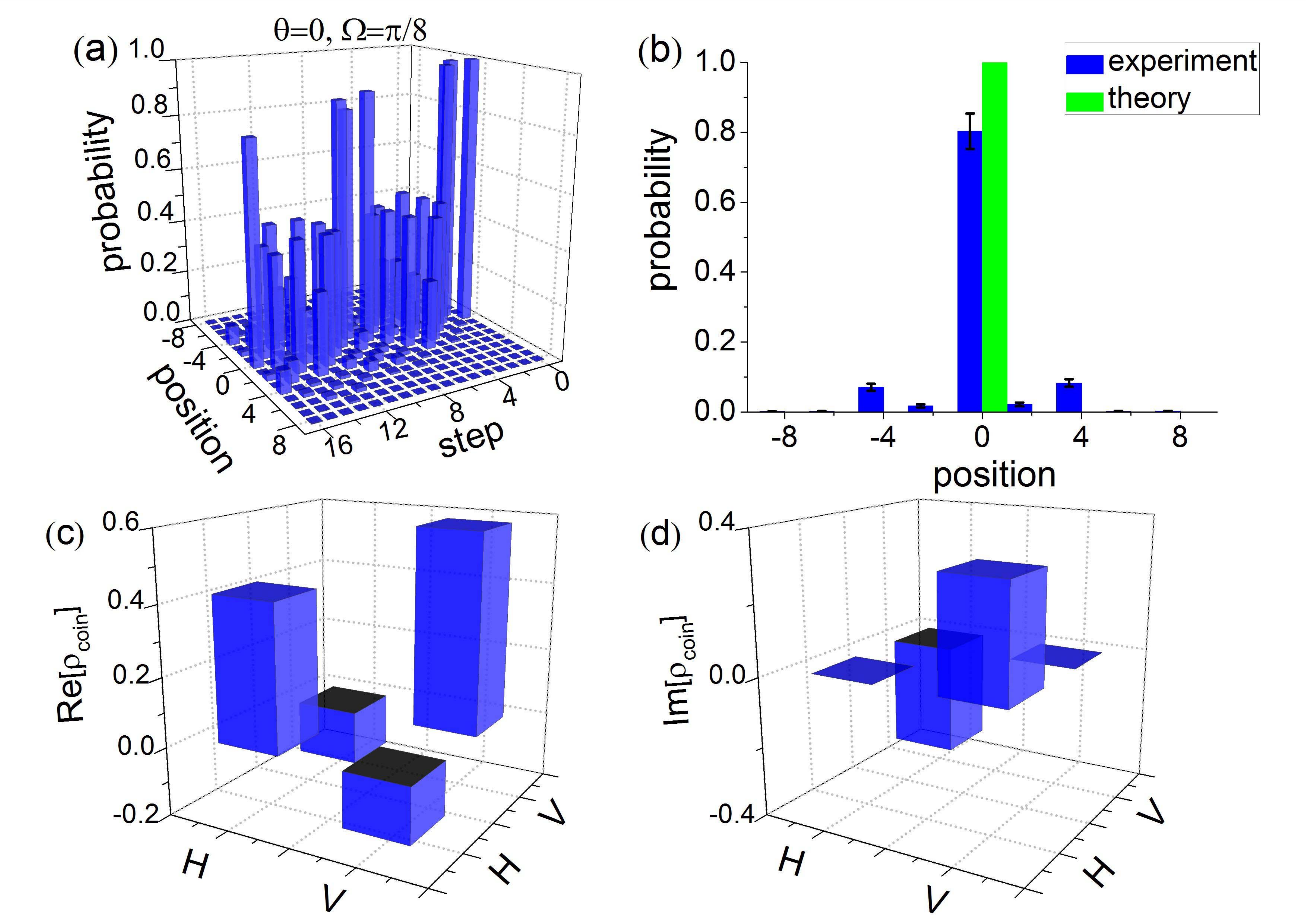}
\caption{%
	(Color online)
	Complete revival.
	(a)~Experimentally measured position distributions (blue bars) for successive
	QW steps up to two revivals ($T=8$) for time-dependent coin parameters
	$\theta=0$ and	$\Omega=\pi/8$,
	commencing with a symmetric coin state.
	(b)~Experimental and theoretical walker probability distribution at the second revival time
	with error bars indicating statistical uncertainty.
	(c)~Real part and
	(d)~imaginary part of the experimentally reconstructed density matrix of the coin
	state after~$16$ steps, which has returned close to its
	original state $(\ket{H}+\text{i}\ket{V})/\sqrt{2}$.%
	}
\label{fig:8steptheta0Omegapi8}
\end{figure}

Specifically we have demonstrated two consecutive complete revivals for
the case that $T=8$ and
also performed the experiment for two other parameter choices that lead to incomplete revivals.
Now we study the case $\theta=0$ and $\Omega=\pi/8$ in detail.
The results are shown in Fig.~\ref{fig:8steptheta0Omegapi8},
which show (a)~the walker distribution from the $1^\text{st}$ step to the~$16^\text{th}$ step,
(b)~the theoretical vs experimental walker distribution at the $16^\text{th}$ step,
and the (c)~real and (d)~imaginary parts of the topographically reconstructed coin state at the~$16^\text{th}$ step.

The walker's distribution exhibits two strong
revivals (Fig.~\ref{fig:8steptheta0Omegapi8})
at the~$8^\text{th}$ and~$16^\text{th}$ steps a pronounced walker-position-distribution peak reforms at the origin.
Specifically we obtain the experimental results
$p^\text{w}_{x=0}=0.918\pm0.037$ and $0.803\pm0.051$ for the first and second revivals with correspondingly low probabilities at the other positions.
From the relation between~$d_\text{TV}$ and $p_0$~(\ref{eq:dTVp0}),
we obtain small $d_\text{TV}=0.082\pm 0.014$ and $0.197\pm 0.043$
for the first and second revivals, respectively.

Additionally we perform a tomographic reconstruction of the coin state
to demonstrate a complete revival of the walker-coin state.
The resultant overlaps
between the reduced coin state and the initial coin state are
$\mathcal{O}=0.965\pm 0.031$ and $0.818\pm 0.024$ for the first and second revivals which equal the overlaps $\mathcal{O}'$ as required for this complete revival.

\begin{figure}
\includegraphics[width=\columnwidth]{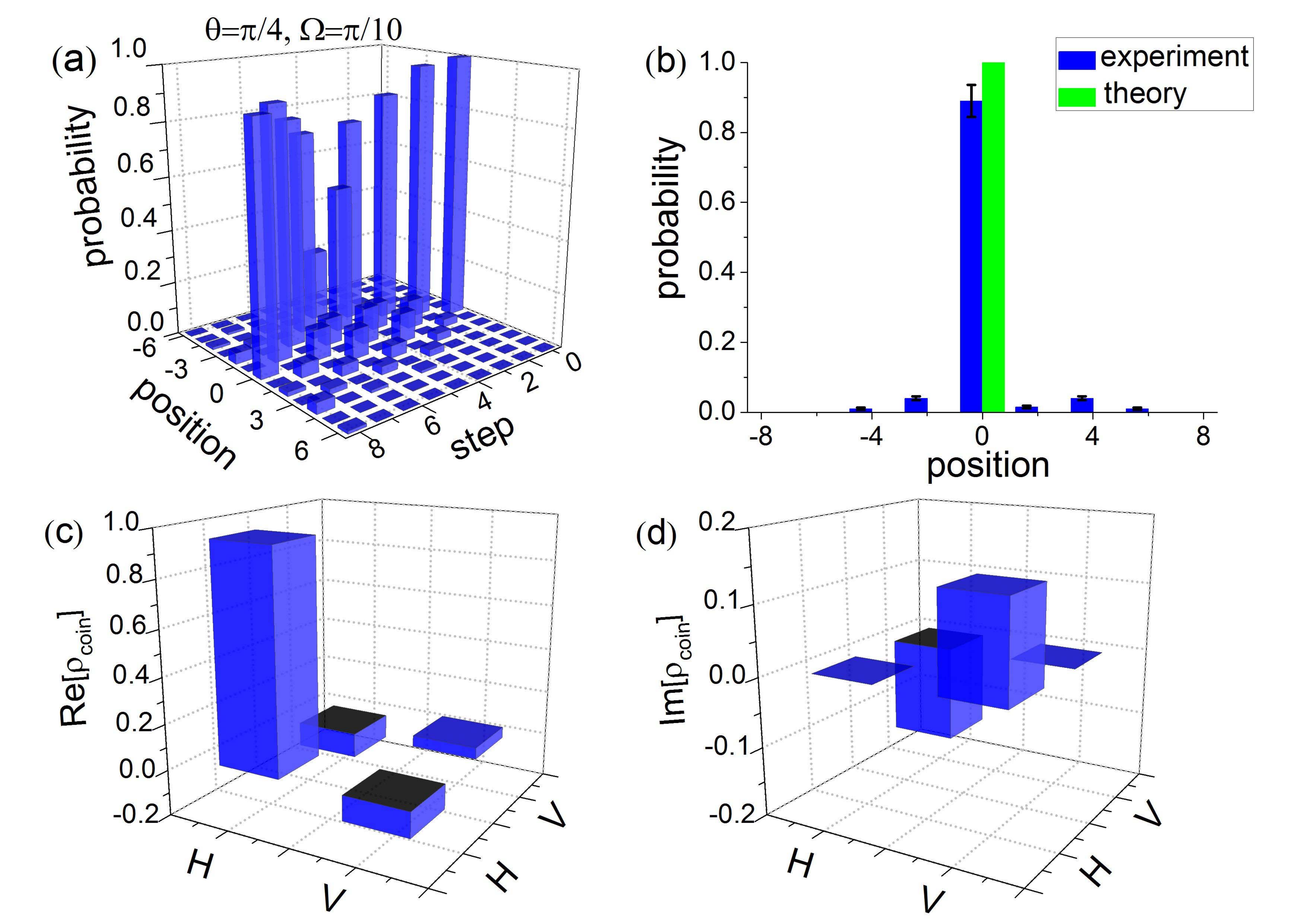}
\caption{%
	(Color online)
	Incomplete revival.
	Same as for Fig.~\ref{fig:8steptheta0Omegapi8}
	except $\theta=\pi/4$ and $\Omega=\pi/10$,
	and the theoretical $T=8$ coin state is
	$0.988\ket{H}+0.156\text{i}\ket{V}$.%
    }
\label{fig:8stepthetapi4Omegapi10}
\end{figure}

To explain the incomplete revival,
we show in Fig.~\ref{fig:8stepthetapi4Omegapi10}
the detailed reduced walker and coin distributions
for $\theta=\pi/4$ and $\Omega=\pi/10$ corresponding to an incomplete revival.
In contrast to the complete revival shown in Fig.~\ref{fig:8steptheta0Omegapi8},
we obtain a low overlap $\mathcal{O}=0.640\pm 0.025$;
this low overlap can be fully understood by realizing that the final coin state is
theoretically predicted to be
$0.988\ket{H}+0.156\text{i}\ket{V}$.
The low~$\mathcal{O}$ for these parameters is merely indicating that the coin state has been rotated from its initial value.
The walker distribution returns to a highly localized state as expected and is,
in that sense,
analogous with the case presented in Fig.~\ref{fig:8steptheta0Omegapi8}.

In contrast to the revival achieved by an evolution followed by its time-reversal,
$U^\dagger U=\one$~\cite{KFC+09},
our experiment demonstrates a controlled-evolution result for
a QW with a time-dependent coin
and specifically shows that we have achieved sufficient control
to realize periodic dynamics of the QW.
We emphasize that our implementation features easy
and convenient adjustability of the coin bias.
Performance of our setup is limited only by imperfections of the
optical components such as nonplanar optical surfaces and
coherence length of single photons, resulting in decoherence, which causes systematic errors in our setup:
the imperfection coherence visibility of the BDI.

Here we have reported incomplete and two complete revivals for effective
time-dependent coin control up to sixteen steps,
but our scheme can be scaled up to realize even more time steps.
The number of sequential BDs grows linearly with the number of time steps,
and the aperture of the last BD also increases linearly with the number of steps
(if the distribution of the walker spreads).
Furthermore the BD surface needs to be flat with high quality.
Thus, scaling up the interferometer is challenging but satisfies the quantum information
scalability principle of carrying an sub-exponential resource overhead.

In summary, we have implemented a stable and efficient way to realize a
one-dimensional photonic QW with time-dependent coin flipping and
thereby observe two revivals of a QW. The time-dependent coin flipping
is divided into two successive non-commuting rotations on the coin,
one of which is time-dependent. Our experiment benefits from the
high stability and full control of both coin and walker at each step
and in each given position.

\acknowledgements
We thank Wei Yi and Xiwang Luo for stimulating discussions.
This work has been supported by NSFC under 11174052 and
11474049, the Open Fund from the SKLPS of ECNU, the 973 Program
under 2011CB921203, the China 1000 Talents program, and Alberta
Innovates Technology Futures.

\bibliography{../qw}

\end{document}